\documentclass{vldb}
\usepackage{amsmath}
\usepackage{mathptmx}

\usepackage{graphicx}
\usepackage{balance}  
\usepackage{url}
\usepackage{subfigure}
\usepackage{enumitem}

\newcommand{\remove}[1]{}

\usepackage[usenames,dvipsnames]{color}

\begin{document}




\title{Continuous Partial Quorums for Consistency-Latency Tuning in Distributed NoSQL Storage Systems}
\subtitle{\textbf{\large{[Please refer to the proceedings of SCDM'15 for the extended version of this manuscript.]}}}

\numberofauthors{3}
\author{
\alignauthor
Marlon McKenzie \titlenote{Authors supported by research funding from Hewlett-Packard Labs, Google, and the Natural Sciences and Engineering Research Council (NSERC) of Canada.} \\
       \affaddr{University of Waterloo, Canada} 
       \email{m2mckenzie@uwaterloo.ca}
\alignauthor
Hua Fan \textsuperscript{\textsuperscript{\normalsize*}}\\
       \affaddr{University of Waterloo, Canada} 
       \email{h27fan@uwaterloo.ca}
\alignauthor
Wojciech Golab \textsuperscript{\textsuperscript{\normalsize*}} \\
       \affaddr{University of Waterloo, Canada} 
       \email{wgolab@uwaterloo.ca}
}



\maketitle

\begin{abstract}
NoSQL storage systems are used extensively by web applications
	and provide an attractive alternative to conventional databases
	when the need for scalability outweighs the need for transactions.
Several of these systems provide quorum-based replication and present
	the application developer with a choice of multiple client-side ``consistency levels''
	that determine the number of replicas accessed by reads and writes, which
	in turn affects both latency and the consistency observed by the client application.
Since using a fixed combination of read and write consistency levels for a given application
	provides only a limited number of discrete options,
	we investigate techniques that allow more fine-grained tuning of the consistency-latency trade-off,
	as may be required to support consistency-based service level agreements (SLAs).
We propose a novel technique called \emph{continuous partial quorums} (CPQ) that assigns the consistency level
	on a per-operation basis by choosing randomly between two options, such as eventual and strong consistency,
	with a tunable probability.
We evaluate our technique experimentally using Apache Cassandra and demonstrate that
	it outperforms an alternative tuning technique that delays operations artificially at clients.
\end{abstract}


\section{Introduction} \label{sec:intro}
NoSQL storage systems are used extensively by web applications
	and provide an attractive alternative to conventional databases
	when the need for scalability outweighs the need for transactions.
Several of these systems, most notably Cassandra \cite{laksh:cass}, Voldemort and Riak,
	are derivatives of Amazon's Dynamo \cite{decandia2007dynamo} and share a common quorum-based
	replication model that enables different behaviors with respect to Brewer's CAP principle,
	which states that during a network partition a system must compromise either consistency or availability \cite{brew:cap}. 
Application developers who use such systems face a choice of multiple client-side ``consistency levels''
	that determine the size of a partial quorum for reads and writes,
	which is the number of replicas that must respond to a read or write request.
This parameter directly affects the latency of read and write operations, and
	indirectly affects the consistency observed by client applications.
Overlapping (e.g., majority) quorums are used to achieve so-called ``strong consistency,''
	meaning that reads always return the latest value of a data object,
	whereas non-overlapping partial quorums provide weaker forms of consistency,
	particularly eventual consistency, whereby reads may return stale values for some period of time after
	an update while replicas of the data object converge to a common state.
In this context a value is \emph{stale} if it has been overwritten by a newer value,
	and is \emph{fresh} otherwise; staleness is a very different concept from the \emph{age} of a value
	with respect to the time it was written into the storage system.

In this paper we investigate the possibility of tuning the consistency-latency trade-off
	in a more fine-grained manner than is possible using client-side consistency levels,
	which offer a limited number of discrete choices (e.g., read one replica, read majority, etc).
Specifically, we focus on techniques that enable fine-grained consistency-latency tuning
	in quorum-replicated storage systems by varying a real-valued parameter, as opposed to the use of 
	a fixed consistency level that offers only a limited number of discrete choices.
Attaining fine-grained control over consistency and latency is an important
	step on the path to supporting service level agreements (SLAs),
	for example where a client application requests that read operations have 95th \%-ile latency 
	at most $L$ milliseconds and return stale values at most $X$ fraction of the time for some thresholds $L > 0$ and $X < 1$.
In this framework a latency-favoring application (e.g., a shopping cart) may specify a lower $L$ and higher $X$,
	whereas a consistency-favoring application (e.g., personal cloud file system)	 may opt for a higher $L$ and lower $X$.
Naturally such SLAs can also specify guarantees on throughput.

Our main technical contribution in connection with fine-grained consistency-latency tuning
	is a novel technique called \emph{continuous partial quorums} (CPQ),
	which entails making a random choice between multiple discrete consistency levels on a per-operation basis.
For example, the application may choose consistency level one with probability $p$
	and majority quorums with probability $1-p$.
In this case $p$ itself becomes a continuous tunable parameter in the range $0 \leq p \leq 1$.
In contrast, using fixed consistency levels for reads and writes and a replication factor of three,
	there are only three possible partial quorums---one, two/quorum, and three/all---%
	and hence nine discrete combinations.
Furthermore, only four of these combinations, namely ones using the one and two/quorum consistency levels,
	provide availability in the presence of a single server failure.

We compare continuous partial quorums experimentally against an alternative technique called \emph{artificial delays} (AD),
	in which clients use a weak consistency level such as read/write one (i.e., operations terminate when one replica responds)
	and boost consistency by injecting a tunable delay immediately before or after executing
	an operation against the storage system.
For example, during a read operation the delay is injected immediately \emph{before} the client
	issues a read request to the storage system.
In this scenario the value returned by the read is fresh as long as it was the last updated value
	at any point in time during the interval starting immediately before the artificial delay
	and ending when the storage system returns a response to the client.
The longer the delay the larger the latency and the higher the odds that
	the read returns a fresh value.

\remove{
The second technique entails tuning protocol timeouts, and is intended to bound latency
	when large partial quorums are used, such as majority quorums or consistency level ``all''
	(i.e., operations terminate when all replicas have responded),
	at the cost of a possibly higher staleness.
In theory such timeouts can be used to control the tail of the latency distribution
	(e.g., 95th and 99th \%-ile latencies).
Timeouts can be controlled separately for read and write operations in recent versions of Cassandra,
	albeit only using a global setting, in contrast to the consistency level, which
	can be set on a per-operation basis.
}

Our experimental comparison of continuous partial quorums against artificial delays
	using Apache Cassandra 
shows that the CPQ technique enables a superior consistency-latency trade-off.
In some cases our technique attains the same degree of consistency (defined more precisely in Section~\ref{sec:meth})
	as artificial delays with severalfold lower latency.


\section{Methodology} \label{sec:meth}

We study the consistency-latency trade-off experimentally by applying the two techniques
	described in Section~\ref{sec:intro} (CPQ and AD) to an Apache Cassandra \cite{laksh:cass}
	cluster deployed in Amazon's EC2 environment.
All EC2 instances are provisioned in the same availability zone and we do not consider 
	geo-replication in this paper.
The workload is generated using the Yahoo Cloud Serving Benchmark (YCSB) \cite{ycsb},
	with a modified Cassandra connector to support our CPQ technique.
YCSB collects precise measurements of throughput and latency.
To measure consistency precisely we follow the approach of Golab et al.\ by calculating
	the $\Gamma$ (Gamma) consistency metric from traces of operations recorded by
	instrumenting YCSB \cite{gamma}.

The $\Gamma$ metric quantifies consistency by measuring how far the behavior of a storage system,
	as observed by client applications (in this case YCSB) deviates from the gold standard of \emph{linearizability}---the
	property that every operation appears to take effect instantaneously at some point between its 
	start time and its finish time.
A $\Gamma$ value of zero for a particular trace of operations indicates linearizable behavior, and positive values
	indicate deviations from linearizability, which we refer to as \emph{consistency anomalies}.
Intuitively, if the $\Gamma$ value is $X > 0$ time units then this indicates that each operation appears to take effect instantaneously
	at some point between its 	start time minus $X/2$ and its finish time plus $X/2$.
Similarly to \cite{gamma}, we calculate a fine-grained form of the metric called the \emph{per-value $\Gamma$ score},
	which quantifies consistency anomalies associated with a collection of operations
	that access the \emph{same key} and read or write the \emph{same value}. 	
Positive $\Gamma$ scores represent an upper bound on the staleness of values returned by read operations.
We use the proportion of positive $\Gamma$ scores as an estimate of the fraction of stale reads,
	which was denoted by $X$ in our discussion of SLAs in Section~\ref{sec:intro}.


Our chosen method of measuring consistency is \emph{client-centric} in the following sense:
	positive $\Gamma$ scores represent consistency anomalies that are actually observed by a collection of clients
	via the responses of read operations.
It is possible for the storage system to contain stale copies of a data item internally even when 
	the $\Gamma$ score is zero, indicating linearizability, as long as the stale copies are never read by clients.
We believe that this approach, which separates the consistency metric cleanly from the implementation details of a storage system,
	is well matched to the task of specifying and verifying SLAs for consistency.

\break
\section{Experiments} \label{sec:exp}

\subsection{Overview}

\subsubsection{Hardware and software environment}
The experiments are staged using six on-demand instances in Amazon EC2, us-west-2b availability zone.
Each host is an m3.2xlarge on-demand instance with 8 virtual
Intel Xeon E5-2670 2.50GHz cores, 32 GB RAM, 2x80 GB SSD local storage. 
The RTT between nodes is 300-450$\mu$s.
Clocks are synchronized to within 2ms using NTP. 
The software environment includes an Ubuntu 14.04 x86\_64 image with Linux kernel version 3.13.0 in
HVM (Hardware Virtual Machine) mode, Oracle Java 1.7.0\_72, Apache Cassandra 2.0.10
	and YCSB 0.1.4 modified as explained in Section~\ref{sec:meth}.
Cassandra is configured with default settings and the data directory
	is placed on SSD-based local instance storage.
Each host runs a single YCSB process with 128 client threads
	that connect to the local Cassandra server.

\subsubsection{Workload and system parameters}
Each experiment comprises a YCSB load phase starting with an empty keyspace,
	followed by a 60-second YCSB transaction phase.
We use a mixture of 80/20\% read/write operations that access 128-byte values.
Keys are generated using one of two YCSB probability distributions, similarly to \cite{gamma}:
	``latest'' with a key space of 1k, and ``hotspot'' with a key space of size 10k and
	80\% of the operations acting on a 20\% hot set of keys.
The replication factor is three.
The target throughput in YCSB is set to 5kops/s/host, and is achieved to approximately 1\%
	in all experimental runs.

\subsubsection{Visualizations}
We present several types of graphs in this section.
Part~(a) of Figures~\ref{fig:sens_quorum},~\ref{fig:sens_ppq} and \ref{fig:sens_delay} presents the proportion of positive $\Gamma$ scores.
%
%
The proportions shown exclude $\Gamma$ scores that are positive but less than the
	clock synchronization threshold of approximately 2ms.
Such small scores do not reliably indicate 	consistency anomalies and we remove them to de-noise our figures.
Part~(b) of Figures~\ref{fig:sens_quorum},~\ref{fig:sens_ppq} and \ref{fig:sens_delay} 
	present the 95\%-ile latencies (ms) corresponding to the runs shown in part~(a),
	calculated as an average over a sample of values reported by YCSB, with one value from each host.

In the interest of readability we do not include error bars in our graphs, but we do observe
	moderate variations in the results.
In particular, the proportion of positive $\Gamma$ scores varies noticeably between runs.
This is partly due to imperfect clock synchronization, which adds noise to measurements of $\Gamma$,
	and partly a side-effect of poor performance isolation in the EC2 environment.
The latency measurements are generally more stable than the consistency measurements, and
	the standard deviation of the 95\%-ile latency reported by YCSB processes
	at different hosts is approximately 1ms.

\remove{
\begin{figure}
\vspace{1in}
\begin{verbatim}
           >> add missing picture <<
\end{verbatim}
\vspace{1in}
\caption{Illustration of per-value $\Gamma$ score. \label{fig:gamma}}
\end{figure}
}

\begin{figure*}[htbp]
  \subfigure[consistency]
    {\includegraphics[width=.5\linewidth]{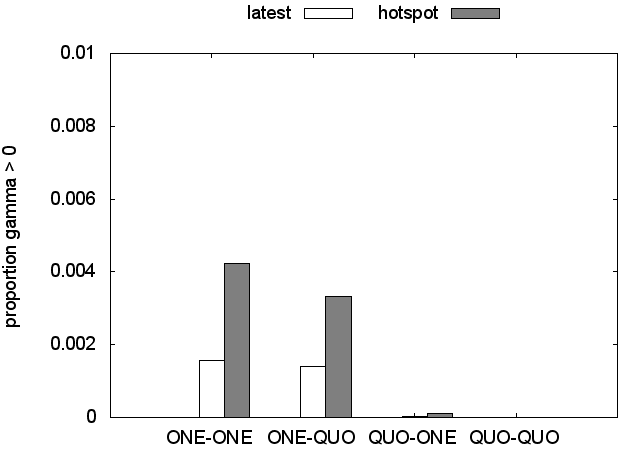}}\hfill
  \subfigure[latency]
    {\includegraphics[width=.5\linewidth]{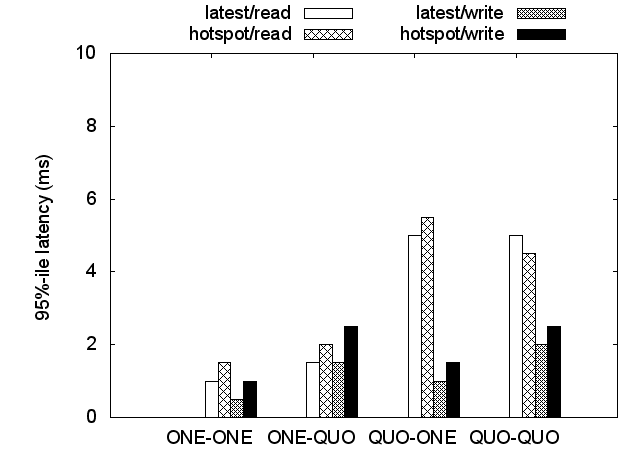}}
  \caption{Consistency and latency vs.\ client-side consistency level (e.g., ONE-QUO means read one, write majority quorum).}
  \label{fig:sens_quorum}
\end{figure*}

\begin{figure*}[htbp]
  \subfigure[consistency]
    {\includegraphics[width=.5\linewidth]{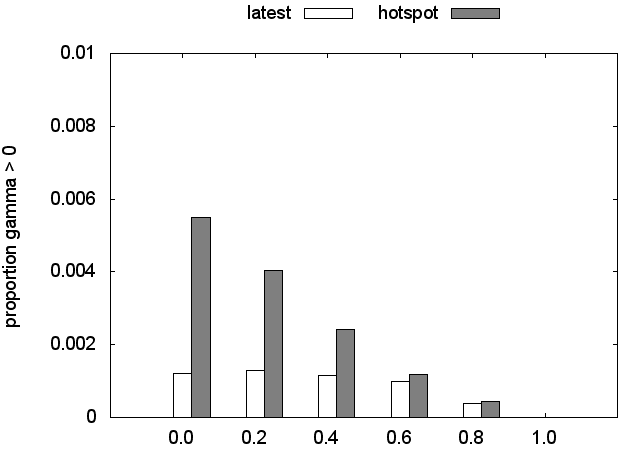}}\hfill
  \subfigure[latency]
    {\includegraphics[width=.5\linewidth]{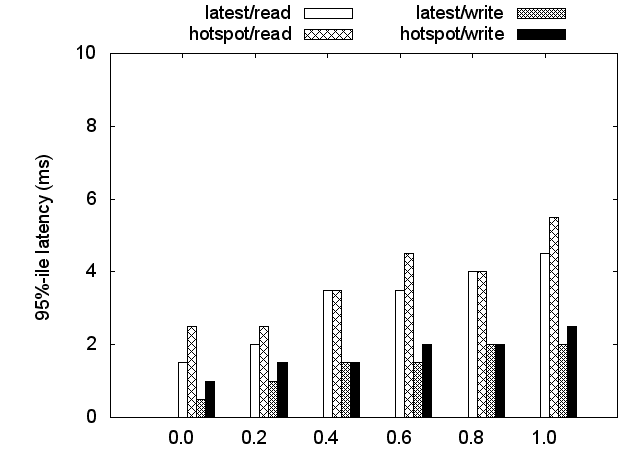}}
  \caption{Consistency and latency versus probability of client-side consistency level quorum vs.\ one.}
  \label{fig:sens_ppq}
\end{figure*}

\begin{figure*}[htbp]
  \subfigure[consistency]
    {\includegraphics[width=.5\linewidth]{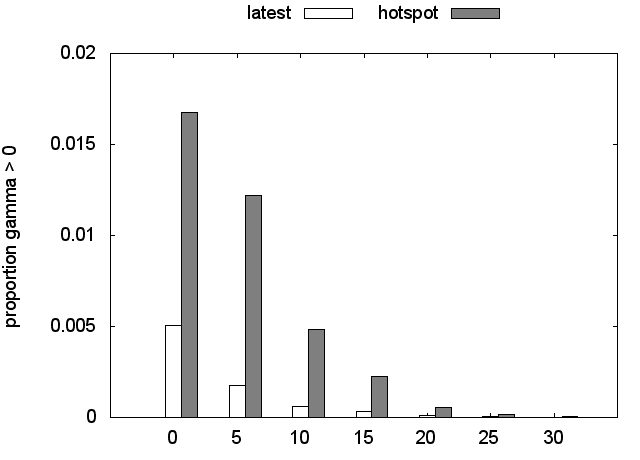}}\hfill
  \subfigure[latency]
    {\includegraphics[width=.5\linewidth]{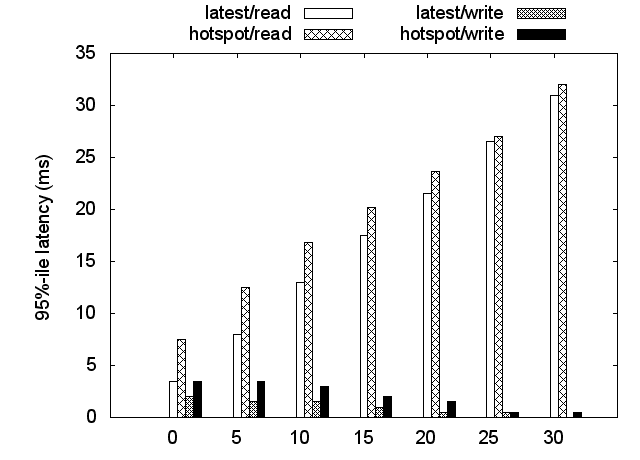}}
  \caption{Consistency and latency versus client-side artificial delay (ms).}
  \label{fig:sens_delay}
\end{figure*}

\subsection{Results}
As a starting point we evaluate the consistency-latency envelope of fixed 
	client-side consistency levels, which is our baseline technique.
We focus specifically on different combinations of one and majority quorum consistency levels,
	which provide availability in the presence of one failed server given the replication factor of three. 	
Figure~\ref{fig:sens_quorum} presents the results.
The x-axis labels are of the form A-B where A and B indicate the client-side
	consistency level for reads and writes, respectively.
Similarly to Figure~6 of \cite{gamma}, our results show that the quorum consistency level
	improves consistency (i.e., lowers the proportion of positive $\Gamma$ scores)
	at the cost of increased latency.
The 95\%-ile latencies are generally less than 8ms, and slightly higher for reads overall
	than for writes---an expected outcome given that Cassandra is write-optimized.
The QUO-QUO case (strong consistency) indicates zero positive $\Gamma$ scores, meaning that the
	storage system produced a linearizable trace.	
In comparison, the ONE-ONE case (eventual consistency) exhibits latencies less than half
	of QUO-QUO, with fewer than 1\% of reads returning stale values.
	
The second set of results, presented in Figure~\ref{fig:sens_ppq}, demonstrates continuous partial quorums in action.
In this experiment the client chooses majority quorum consistency with probability $p$, shown on the x-axis,
	and one consistency with probability $1-p$.
The same policy is used for both read and writes.
As $p$ increases from 0 to 1 we observe that both the consistency and latency
	gradually morph from values corresponding to the ONE-ONE case in Figure~\ref{fig:sens_quorum}
	to values corresponding to the QUO-QUO case.
Thus, CPQ successfully attains points in the two-dimensional consistency-latency 
	spectrum that lie in-between the discrete points attained using fixed client-side consistency levels.
Furthermore, when $p$ is chosen between 0 and 1, CPQ attains trade-offs that are not possible
	at all using fixed client-side consistency levels.
In particular, these points do not correspond to the ONE-QUO and QUO-ONE cases in Figure~\ref{fig:sens_quorum}.
Aside from differences in the $\Gamma$ proportion, these configurations provide a different balance of read and write
	latencies compared to CPQ.

The last set of results, presented in Figure~\ref{fig:sens_delay}, demonstrate the behavior of artificial delays.
In this experiment the client uses consistency level one for both reads and writes, and
	boosts consistency by injecting a delay at the beginning of each read.
The length of the delay in milliseconds is shown on the x-axis, and contributes directly to the latency of read operations.
For example, with a 20ms delay the 95\%-ile latency for reads is 20-25ms, compared to 1-3ms in Figure~\ref{fig:sens_quorum} (ONE-ONE case)
	and Figure~\ref{fig:sens_ppq} (0.0 case).
(Note that the consistency and latency scales in Figures~\ref{fig:sens_quorum} and \ref{fig:sens_ppq} range from 
	0 to 0.01 and 0 to 10ms, respectively, whereas in Figure~\ref{fig:sens_delay} they range from 0 to 0.02 and 0 to 35ms, respectively.)
At this point the $\Gamma$ proportion approaches zero, which is the value attained using majority quorums
	in Figures~\ref{fig:sens_quorum} (QUO-QUO case) and \ref{fig:sens_ppq} (1.0 case) at a latency of only 4-6ms for reads
	and 2-4ms for writes.
Thus, a 20ms artificial delay achieves slightly worse consistency than quorum operations with severalfold higher latency.
Even with a 5ms delay the read latency in Figure~\ref{fig:sens_delay} exceeds that of quorum reads,
	but the consistency observed is only slightly better than using consistency level one and no delay.
Thus, artificial delays provide a suboptimal consistency-latency trade-off compared
	to both our CPQ technique and the baseline technique.

\section{Related Work} \label{sec:rel}
Recent research in the area of consistency has addressed
	the classification of consistency models, consistency measurement,
	and the design of storage systems that provide precise consistency guarantees.
This body of work is influenced profoundly by the CAP principle,
	which states that a distributed storage system must make a trade-off
	between consistency (C) and availability (A) in the presence of a network partition (P) \cite{brew:cap}.
The PACELC formulation builds on CAP by considering two separate cases:
	during a network partition it reduces directly to CAP, but
	during failure-free operation it dictates a trade-off between latency and consistency \cite{abadi:cap}.

Distributed storage systems use a variety of designs that achieve different trade-offs with respect to CAP.
Amazon's Dynamo and its derivatives (Cassandra, Voldemort and Riak) use a quorum-based replication scheme
	that can operate either in CP (i.e., strongly consistent but sacrificing availability) or AP (i.e., highly available but eventually consistent)
	mode depending on the size of the partial quorum used to execute read and writes \cite{laksh:cass, decandia2007dynamo}.
The techniques discussed in this paper--CPQ and AD---are targeted specifically at this family of systems.
Since they are implemented at clients these techniques can be used with any quorum-replicated
	system that supports tunable partial quorums.

Many alternative designs have been proposed for supporting stronger notions of consistency in storage systems.
Bigtable provides atomic access to individual rows, and is eventually consistent when deployed across multiple data centers \cite{DBLP:journals/tocs/ChangDGHWBCFG08}.
PNUTS provides per-record timeline consistency, which ensures that replicas of a record apply updates in the same order \cite{DBLP:journals/pvldb/CooperRSSBJPWY08}.
COPS provides causal consistency with convergent conflict handling and read-only transactions, and is designed for wide-area deployments \cite{cops:sosp11}.
Causal consistency is in some sense the strongest property that can be guaranteed in the presence of network partitions, which makes COPS an AP system in the context of CAP \cite{mahajan2011consistency}.
Bolt-on causal consistency is a shim layer that provides causal consistency on top of eventual consistency \cite{Bailis:2013:BCC:2463676.2465279}.
Spanner is a geo-replicated transactional database that provides external consistency, which is similar in spirit to Lamport's atomicity property (see Section~\ref{sec:meth}) \cite{goo:span}.
The replication and transaction commitment protocols in these systems are geared toward specific notions of stronger-than-eventual consistency
	and do not expose a client-side consistency level setting that could be used with our CPQ technique.

\remove{
Most distributed datastores currently being used
	\cite{laksh:cass,cops:sosp11,vogels:dynamo,DBLP:journals/tocs/ChangDGHWBCFG08,DBLP:journals/pvldb/CooperRSSBJPWY08,goo:span}.
 	make a choice of a weaker level of consistency that
	the applications using it would be tolerant to. The applications will see a version of the data which may or
	may not be the most recent at that point, but after some period of time replicas are eventually brought
	up to date with the most recent set of changes. This is the model of eventual Eventual Consistency, and this
	model is used in combination with the tradeoffs discussed to provide a model of the storage suitable for a specific
	application requirement. While this model is suitable for some applications, a lot of applications have
	more stringent consistency requirements on its data, or the data may have various clients which different
	consistency or latency requirements, creating the neeed for more client aware services.
}

Several systems consider the problem of providing continuously tunable consistency guarantees.
TACT is a middleware layer that uses three metrics to express consistency requirements with respect to read and write operations: numerical error, order error, and staleness \cite{conit}.
TACT relies on a consistency manager that pushes updates synchronously to other replicas.
Pileus allows client applications to declare consistency and latency requirements in the form of SLAs \cite{terry2013consistency}.
These SLAs include latency and staleness bounds but do not support the types of probabilistic guarantees discussed in Section~\ref{sec:intro}.
Internally, Pileus enforces the SLAs by choosing which replica to access in an SLA-aware manner, whereas Dynamo-style systems tend
	to always access the closest replicas.
Tuba supports consistency SLAs by automatically reconfiguring the locations of its replicas in response to the client's location and request rates \cite{ardekani2014self}.
AQuA is middleware layer that allows the client application to specify latency	and consistency requirements similarly to Pileus, but with a focus on
	time-sensitive applications \cite{ksc:qos}.
It provides probabilistic timeliness guarantees by selecting replicas dynamically using probabilistic models.

We are aware of only two systems that use artificial delays for consistency-latency tuning.
Golab and Wylie propose consistency amplification---a framework for supporting consistency-based
SLAs by injecting client-side or server-side delays whose duration is determined adaptively using
measurements of the consistency actually observed by clients \cite{conamp}.
Rahman et al.\ present a similar system called PCAP, where delays are injected only at clients and their duration is determined using
	a feedback control mechanism \cite{pcap}.
PCAP also varies the read repair rate, which is shown to be a far less effective tuning knob.
The evaluation of the system considers the proportion of operations that satisfy
	particular consistency and latency requirements, and does not investigate the optimality of this trade-off
	with respect to fixed client-side consistency levels such as majority quorums.
The argument given against strict quorums is that they may cause storage operations to block
	in the event of a network partition.
However, the consistency calculations used to tune artificial delays in PCAP are themselves
	blocking because they are based upon operation logs collected from multiple servers.
Furthermore, in practice even quorum operations can be made non-blocking by using read and write timeouts,
	which are configurable in recent versions of Cassandra.
Timeouts ensure that every operation eventually either completes successfully, or fails
	and allows the client to retry the operation using a smaller partial quorum.

The use of server-side artificial delays is explored in \cite{fan} as a technique for reducing the severity of consistency
	anomalies in Cassandra when client-side consistency level one is used.
The delays are injected judiciously following the garbage collection stop-the-world pause, which improves consistency
	drastically with negligible impact on latency.
In contrast, the artificial delays used in PCAP and explored in our own experiments
	incur a latency penalty for every single read operation, which increases average latency directly.
		

\remove{	
Among these two tuning knobs 
relies predominanelt
 is both a principle
	(based on the original CAP principle), and a system that provides adaptable client requirements based on
	availability and/or consistency service level requirements. The system extends the CAP principle (which
	makes a choice between consistency and availability in the presence of a network partition) to be a tradeoff
	between all three parameters, consistency (C), availibility (A), partitioning (P), with each being probabilistically
	modelled. The system gives the client the ability to specify a consistency or availability Service Level Agreement
	(SLA), and it adapts, with the changing network conditions, to satisfy each SLA. 
}
		
In the pursuit of an empirical understanding of CAP-related trade-offs
	several papers have explored techniques for measuring consistency
	 \cite{gamma,wada:cidr,Bermbach,zellag2012consistent}.
Measuring consistency in a precise way is subtly difficult because consistency anomalies
	such as stale reads are the result of interplay between multiple storage operations.
As a result, some of the contributions in this space consider simplified techniques
	that measure the convergence time of the replication protocol rather than
	the consistency actually observed by client applications (e.g., \cite{wada:cidr,Bermbach})
	or quantify the consistency observed in terms of quantities that do not translate directly
	into staleness measures expressed naturally in units of time (e.g., counting cycles in a dependency graph \cite{zellag2012consistent}).
Probabilistically bounded staleness (PBS) is a mathematical model of partial quorums
	that overcomes these limitations but is based upon the simplifying assumption that writes
	do not execute concurrently with other operations \cite{pbs}.
The theory underlying probabilistic quorum systems was originally developed by Malkhi, Reiter, and Wright \cite{pqs}.

\remove{
Work has also been done to provide stronger consistency guarantees using datastores with weaker consistency
	models. Peter Balis et al. in \cite{Bailis:2013:BCC:2463676.2465279} implemented a shim layer which is used
	on top of any eventually consistenty datastore, providing ``Bolt-On'' Causal Consistency to the clients
	using the layer. Their choice of providing Causal Consistency was to provide a stronger consistency level
	that eventual, and due to recent work stating that Causal Consistency is the strongest that can be
	guaranteed in the presence of network partitions \cite{mahajan2011consistency}.
}

\section{Discussion and Conclusion} \label{sec:discuss}

Our experiments using Cassandra in Amazon's EC2 environment demonstrate
	that the consistency-latency trade-off can be tuned in a continuous manner
	using only a handful of discrete client-side consistency levels.
We achieve this goal using a novel technique called continuous partial quorums (CPQ),
	which chooses randomly between two discrete consistency levels
	according to a tunable probability parameter.
Compared to client-side artificial delays with consistency level one,
	CPQ is able to achieve a more attractive consistency-latency trade-off,
	in some cases offering the same degree of consistency
	with severalfold lower latency.
This result confirms informal claims regarding the potentially detrimental 
	effect of injecting artificial delays (e.g., see \cite{pbs}), albeit only
	in the special case where the delay is in the critical path of every read operation.
As discussed in Section~\ref{sec:rel}, delays injected at servers
	can improve consistency effectively with a very small latency penalty \cite{fan}.

Although we demonstrate CPQ specifically in the context of Apache Cassandra,
	the technique is applicable to any system that supports a set of
	discrete client-side consistency options.
In future work we plan to implement and evaluate this technique on top of other storage systems
	and expand the scope of experiments to cover geo-replication.
Furthermore, we plan to construct a comprehensive middleware framework that uses CPQ
	and other tuning techniques to supporting probabilistic consistency and latency guarantees.

\remove{
Indeed our experiments show that quorum reads and writes provide both
	lower latency on average and better consistency than
	eventually consistent operations augmented with delays of 12ms.
To match quorum operations in terms of observed consistency, the delays would need to be comparable
	to the largest $\Gamma$ scores observed in runs without delays.
A quick inspection of our experimental numbers reveals that such delays, and hence the average latencies,
	would loosely correspond to the \emph{maximum} latencies observed using quorum operations (i.e., 100s of ms).
This point in the consistency-latency trade-off	 could make sense if
	quorum operations are either not supported or too costly.
}


\balance
\bibliographystyle{abbrv}
\bibliography{bibfile}

\end{document}